\documentstyle[twoside,fleqn,espcrc2,epsf]{article}

\newcommand{\AmS}{{\protect\the\textfont2
  A\kern-.1667em\lower.5ex\hbox{M}\kern-.125emS}}

\newlength{\numlen}
\newcommand{\n}{\settowidth{\numlen}{0}\makebox[\numlen]{}}

\newcommand{\la}[1]{\label{#1}}

\newcommand{\be}{\begin{equation}}
\newcommand{\ee}{\end{equation}}
\newcommand{\ba}{\begin{eqnarray}}
\newcommand{\ea}{\end{eqnarray}}

\newcommand{\h}{{\hspace{0.5 cm}}}

\newcommand{\tr}{{\rm Tr\,}}
\newcommand{\nr}[1]{(\ref{#1})}
\newcommand{\fr}[2]{{\frac{#1}{#2}}}

\def\lsim{\raise0.3ex\hbox{$<$\kern-0.75em\raise-1.1ex\hbox{$\sim$}}}
\def\gsim{\raise0.3ex\hbox{$>$\kern-0.75em\raise-1.1ex\hbox{$\sim$}}}

\newcommand{\half}{{\scriptstyle{1\over2}}}
\newcommand{\quarter}{{\scriptstyle{1\over4}}}

\newcommand{\etal}{{et al.}}

\newcommand{\fig}[1]{Fig.\,\ref{#1}}
\newcommand{\eq}{eq.\,}

\newcommand{\msbar}{\overline{\mbox{\rm MS}}}

\newcommand{\figtopspace}{\vspace*{-1cm}\hspace{-3mm}}
\newcommand{\figbottomspace}{\vspace*{-5.1cm}}
\newcommand{\figfxsize}{8.6cm}
\newcommand{\phisq}{\langle\phi^\dagger\phi\rangle}
\newcommand{\secspace}{}
\newcommand{\sectopspace}{}

\title{Finite $T$ electroweak phase transition on the lattice%
\footnotemark[2]
}
\author{
	K. Rummukainen\address{Indiana University, Department of Physics,
	Swain Hall-West 117, Bloomington, IN 47405, USA}%
\thanks{\mbox{e-mail: kari@trek.physics.indiana.edu}
        \mbox{~Address after 9/96: Universit\"at Bielefeld, Germany}
	\mbox{$^\dagger$Review talk at LATTICE96, St.~Louis, USA, June 1996}}
\hfill \raisebox{2.5cm}[0cm][0cm]{IUHET-341/August 1996}
}

\begin{document}
\thispagestyle{empty}

\begin{abstract}
This talk reviews recent lattice results on the high $T$ electroweak
phase transition.  A remarkably accurate picture emerges: a) the
transition is of first order for $m_H \lsim 80$\,GeV and vanishes for
larger $m_H$; b) transition temperature, latent heat and interface
tension are known, as well as c) the properties of the broken and
symmetric phases.  New developments in the sphaleron rate calculations
are discussed.
\end{abstract}

\maketitle

\sectopspace
\section{INTRODUCTION}
\secspace

At high temperatures the spontaneously broken symmetry of the
electroweak (EW) sector of the standard model is restored
\cite{Kirzhnitz76}.  How this restoration occurs is crucial for
understanding the effects of EW physics to the baryon number of the
Universe, and for the viability of EW baryogenesis
\cite{baryogenesis}.  The essential questions are: what is the order
of the transition, or is there only a smooth cross-over?  What are the
values of the transition temperature $T_c$, latent heat $L$, interface
tension $\sigma$ and the discontinuity of the Higgs condensate?  Very
important are also the properties of the phases close to $T_c$: the
equation of state, screening lengths, and phase metastability ranges.
For baryogenesis scenarios, a crucial quantity is the sphaleron rate,
i.e. the rate of the baryon number fluctuation.  These quantities are
parametrized by the still unknown Higgs mass $m_H$ (experimentally
$m_H \gsim 64$\,GeV).  All of the questions above have been addressed
with lattice simulations; and happily, more often than not,
definite answers have been found.

Given the success of the EW perturbation theory at $T=0$, it is
natural to apply it at finite $T$: indeed, the effective potential has
been calculated up to 2-loop level \cite{2loop,fkrs_framework}.
However, it has become clear that the perturbation theory in gauge
theories at high $T$ is intrinsically unreliable due to infrared
divergences \cite{Linde80}.  At low $T$ the perturbative expansion is
regulated by the large value of the Higgs condensate $v$: the
expansion parameter is $\sim g^2T/[ \pi m_W]$.  When $T$ increases,
$v$ becomes smaller, and in the symmetric phase the expansion cannot
be controlled even with resummation techniques.  Clearly, a
non-perturbative approach is required.

In comparison with the QCD phase transition, until recently the finite
$T$ EW transition had not been extensively studied.  Since the
non-perturbative effects are expected to be mainly due to the SU(2)
gauge fields, the studies have concentrated on the SU(2) gauge + Higgs
model.  A direct way to study the finite $T$ physics is to perform 4D
Euclidean lattice simulations; the standard formalism was set up
almost a decade ago \cite{Damgaard87,Evertz87}, but the first results
close to the physical weak coupling parameter values were published
only in 1992 \cite{Bunk92}.  The development of the dimensionally
reduced 3 dimensional effective formalism, initiated in \cite{krs1}
and completed in \cite{fkrs_framework,klrs_generic}, was an important
milestone theoretically and especially for practical simulations.
Until Lattice '95, Higgs mass ranges $m_H=18$--49\,GeV had been
investigated with 4D simulations
\cite{desy_18,desy_18_49,desy_35interface} and 35--80\,GeV with the 3D
formalism \cite{krs1,fkrs_80,fklrs,Ilgenfritz95} (for earlier reviews,
see \cite{Jansen96,Kajantie95,Shaposhnikov92}).  In this conference
new 4D results with $m_H \lsim 102$\,GeV were presented by the DESY
group \cite{desy_34,Hein_16,Csikor_asymm_sigma} and Y.\,Aoki
\cite{Aoki96}, and 3D results with $m_H \le 180$\,GeV by Kajantie
\etal \cite{klrs_nonpert,klrs_heavy}, Karsch \etal
\cite{Karsch_3d80,Karsch_3d100}, G\"urtler \etal \cite{Gurtler96} and
Philipsen \etal \cite{Philipsen96}.  New studies of the sphaleron rate
were reported by Tang and Smit \cite{Tang96}.

\sectopspace
\section{THE SU(2)-HIGGS MODEL}
\secspace

Since the essential physics of the EW phase transition is expected to
arise from SU(2) gauge fields and the Higgs field, let us for the
moment neglect  SU(3) and U(1) components of the gauge fields and the
fermions entirely.  The SU(2)-Higgs Euclidean Lagrangian is
\ba
  L &=& \quarter F_{\mu\nu}^a F_{\mu\nu}^a + 
	(D_\mu\phi)^\dagger(D_\mu\phi)  \nonumber \\
   & - & m^2 \phi^\dagger\phi + 
        \lambda (\phi^\dagger\phi)^2  \,, 
\la{4dlagrangian}
\ea
where $\phi = (\phi^+,\phi^0)$ is the Higgs doublet.  In the
following we discuss the derivation of the effective 3D action
from \nr{4dlagrangian}.

\sectopspace
\subsection{3D effective action: why and how}\label{sec:3daction}

Because of the relatively small value of the coupling $g$,
the EW theory at high temperatures has a very wide range of mass
scales ($\sim$ inverse screening lengths):
\be
  T \gg m_D \approx gT \gg g^2 T,\,m_W(T),\,m_H(T). \la{scales}
\ee
For example, the simulations have shown that if $m_H\approx 60$\,GeV,
then $m_H(T_c) < 0.1 T_c$.  On an Euclidean finite $T$ system the
imaginary time is restricted to the interval $0\le\tau\le 1/T$; thus,
when this system is latticized, in order to avoid finite size effects
the ratio of the spatial and temporal extents of the lattice should be
at least $L_s/L_\tau \gsim 30$--40.  This is a very punishing
requirement.

The extreme `flatness' of the geometry already suggests that the
essential features of the system might be described by an effective 3D
theory.  Because of the periodic boundary conditions, when the bosonic
fields are expanded in Fourier modes the inverse propagator becomes
$[\vec k^2 + m_0^2 + (2\pi n T)^2]$.  If $T$ is large compared to the
other relevant mass scales of the system, the non-static Matsubara
modes with $n = \pm 1,\pm 2,\ldots$ acquire a heavy mass $2\pi n T$
and can be integrated over.  What remains is an effective 3D theory of
the static ($n=0$) modes.  If the original 4D theory has fermions,
then, because of the antiperiodicity in $\tau$, {\em all} fermionic
modes become massive with $m = (2n+1)\pi T$ and can be integrated
over.  The effect of the fermion fields is only to adjust the
parameters of the effective 3D bosonic theory.  The derivation of the
3D effective theory, {\em dimensional reduction\,} (DR), introduced in
\cite{dimred}, was fully developed for the EW theory in
\cite{fkrs_framework,klrs_generic,Laine95}.

The integration over the non-static modes can be performed {\em
perturbatively}.  This is possible if

$\bullet$ $g^2$, $\lambda \ll 1$

$\bullet$ $T \gg m_Q(T_c)$, the relevant mass scales at $T_c$.

For the EW transition, these conditions are met when $30 \lsim m_H
\lsim 240$\,GeV: the lower bound comes from the requirement that
$m_H(T)$, $m_W(T)$ must be $\ll T$ in the vicinity
of $T_c$, and the upper bound from the fact that for large $m_H$ the
EW theory becomes strongly coupled.  Note that there are no IR
problems in the {\em derivation\,} of the effective action: the IR
modes are inherently 3-dimensional and are not integrated over.
Indeed, the effective theory retains all IR divergences of the
original theory!  Moreover, the perturbative DR does not require that
the finite $T$ perturbation theory should be applicable in general:
the criterion for finite $T$ perturbation theory to work is $g^2 T/m_Q
\ll 1$, which is not satisfied at high $T$.

Starting from eq.~\nr{4dlagrangian} the effective theory can be
defined by the action $S_3[A^a_0(\vec x),A_i^a(\vec x),\phi(\vec x)]$,
a 3D SU(2) gauge + adjoint Higgs + fundamental Higgs theory, where the
coefficients of the action depend on the original 4D action
coefficients.  The adjoint Higgs $A_0$ is the remnant of the timelike
component of the 4D gauge field.  This action was used in
\cite{krs1,fkrs_80} to simulate $m_H=35$ and 80\,GeV systems.
Further, for the EW theory, one observes that the Debye mass $m_D =
\sqrt{5/6}gT$ is large, and the field $A_0$ can also be integrated
over.  The resulting action
\ba
  S_3[A_i,\phi] &=& \int d^3x \big[
  \quarter F^a_{ij}F^a_{ij}+
  (D_i\phi)^{\dagger}(D_i\phi) \nonumber \\
  &+& {m}_3^2\phi^{\dagger}\phi+{\lambda}_3
  (\phi^{\dagger}\phi)^2 \big] \la{3daction}
\ea
has been the `workhorse' in all recent 3D simulations
\cite{klrs_nonpert,Gurtler96,Karsch_3d80}.  It is similar in form to
the original 4D action \nr{4dlagrangian}; however, now the fields and
couplings have dimensions $[\phi] =\mbox{GeV}^{1/2}$, $[g_3^2] =
[\lambda_3] = \mbox{GeV}$.  In \cite{Karsch_phi4} the action was
further simplified to an O(4)-symmetric scalar theory by integrating
over the gauge fields.  However, this action failed to describe the
phase transition correctly, indicating the essential role of the
magnetic sector of the gauge fields.

The theory is now uniquely fixed by three parameters, the gauge
coupling $g_3^2$ and two dimensionless numbers
\be
  x \equiv \lambda_3/g_3^2, \h y \equiv
  m_3^2(g_3^2)/(g_3^2)^2 \la{3dparams}
\ee
The action \nr{3daction} is {\em superrenormalizable:} the couplings
$g_3^2$ and $\lambda_3$ do not run (in $\msbar$), and $m_3^2$ has only
linear and log-divergences at 1- and 2-loop levels.  This property 
makes the relation between lattice action and continuum parameters
\nr{3dparams} particularly transparent, as discussed in section
\ref{sec:3dlatt}.

The action of form \nr{3daction} can be derived already with 1-loop
DR.  However, to fully utilize the superrenormalizability better
accuracy is required, and in
\cite{fkrs_framework,klrs_generic,klrs_nonpert} DR is
performed with {\em Green's function matching\,}: one starts from
a general 3D superrenormalizable action and matches all 2- and 4-point
Green's functions to the static 4D Green's functions of the original
theory.  To cancel large logarithms (in $\msbar$), the 4D couplings
are run to scale $\mu = 4\pi Te^{-\gamma} \approx 7T$ by the standard
4D $\beta$ functions.  The matching is done to a consistent accuracy
in $g^2$, $\lambda$.  The action \nr{3daction} can provide the
relative accuracy $\delta G/G \sim O(g^3)$.  To go beyond this would
require the inclusion of 6-dim.\@ operators while giving up
superrenormalizability.  By calculating the effects of the excluded
higher dimensional operators the accuracy of the effective action can
be estimated \cite{klrs_generic}.  By construction, the Green's
function matching avoids the non-localities inherent in the standard
integration and `block transformation' -type approaches to effective
actions \cite{Jakovac96}.

The DR process provides us with the essential relations between the 3D
and 4D parameters.  There is a large class of 4D theories which map
into the 3D SU(2)+Higgs theory.  Since the fermions do not `survive'
DR, this class includes the 4D theory of SU(2) + Higgs + fermions and
the minimal standard model (MSM), where the effect of U(1) gauge field
can be estimated perturbatively.  The mapping
$(g_3^2,x,y)\leftrightarrow$\,4D parameters for these theories has
been worked out in detail in \cite{klrs_generic}.  Conversely, a
single 3D simulation can yield physical results for the whole class of
4D theories.  Recently DR has been worked out for MSSM by several
authors \cite{mssm}.

For the 4D SU(2)-Higgs theory the relation 4D$\leftrightarrow$3D
is \cite{klrs_generic,klrs_nonpert}
\ba
  g^2_3 &=& 0.44015 T^*   \la{g32} \\
  x & = & -0.00550 + 0.12622 h^2 \la{x} \\
  y & = & 0.39818 + 0.15545 h^2 \nonumber \\
    & - & 0.00190 h^4 - 2.58088 (m_H^{*}/{T^*})^2, \la{y}
\ea
where $h=m_H^*/m_W$, $m_W=80.6$\,GeV and
\be
  g = \fr23, \h \lambda = \fr18 g^2 (m_H^*/m_W)^2
\ee
(The authors of \cite{Karsch_3d80,Gurtler96} use somewhat different
conventions.)  The notation $m_H^*$, $T^*$ is used to remind that
these are not the physical $T$ or the pole $m_H$; for the SU(2)-Higgs
theory without fermions the difference is small.  These parameters are
used extensively to present the results of 3D simulations.

\sectopspace
\subsection{Lattice action in 4D}\label{sec:4daction}

The lattice action is conventionally written as
\ba
  S &=& \beta_G \sum_{x;\,\mu<\nu} (1 - \fr12\tr P_{x;\mu\nu})  \nonumber \\
    &-& \beta_H \sum_{x;\,\mu} 
	\fr12 \tr \Phi^\dagger_x U_{x;\mu}\Phi_{x+\hat\mu}) 
\la{lattaction} \\
    &+& \sum_x \fr12 \tr \Phi^\dagger_x\Phi_x  
	+ \beta_R \sum_x \bigl[ \fr12\tr\Phi^\dagger_x\Phi_x -1 \bigr]^2 
	\nonumber
\ea
where $\Phi = R\times V$, $V\in\mbox{SU(2)}$, and $R^2 =
\half\tr\Phi^\dagger\Phi$.  The Higgs field $\Phi$ has
$\mbox{SU(2)}^{\rm gauge}\otimes\mbox{SU(2)}^{\rm global}$ symmetry.

The essential question is now the relation of the lattice parameters
to continuum physics.  In the 4D formalism this is done 
non-perturbatively, relying only on the measurements of physical
quantities.  This comes at a cost: simulations at $T=0$ are needed in
order to set the scale.  At tree level, the relation is
\ba
  \beta_G &=& 4/g^2 \la{4dlatcoupling} \\ 
   (ma)^2 &=& (1 - 4\beta_H - 2\beta_R)/\beta_H \\
  \lambda &=& 4 \beta_R / \beta_H^2.
\ea
Strictly speaking, for the 4D SU(2)-Higgs the $a=0$ limit cannot be
reached because of triviality.  The term ``continuum limit'' in this
case means reaching a region where the cut-off effects become negligible:
physics remains invariant when moving along the renormalization group
trajectories, or {\em constant physics curves\,} (CPC).

The most detailed scaling study so far has been performed by the DESY
group \cite{desy_34} using lattices with $N_\tau = 1/(aT) = 2$--5\@.
A practical way to determine CPCs is to use the transition
itself as follows \cite{desy_18_49,desy_34}:

\noindent
(a) To be close to the desired physical point $g^2 \approx 0.5$,
$m_H\approx 34$\,GeV, couplings $\beta_G=8$, $\beta_R=0.0003$ are
chosen for $N_\tau=2$ simulations.  $m_W$ is fixed to 80\,GeV\@. The
value of $\beta_H$ is tuned until the transition coupling
$\beta_{H,c}$ is found.

\noindent
(b) Using these couplings one performs $T=0$ ($N_\tau \gsim N_s$)
simulations.  Higgs and W masses and the static potential (Wilson
loops) are measured; from the static potential the renormalized gauge
coupling $g_R^2$ can be extracted.  This also gives $\lambda_R \equiv
g_R^2/8(m_H/m_W)^2$.

\noindent
(c) Using the known continuum 1-loop RG-equations for $\beta_G$ and
$\beta_R$, one runs along CPC from $N_\tau \rightarrow N_\tau+1$
($a\rightarrow a N_\tau/(N_\tau+1)$).
The step (a) is then repeated with the new couplings.

Good scaling now means that dimensionless physical quantities $g_R^2$,
$\lambda_R$, $T_c/m_W$ remain invariant.  This is surprisingly well
satisfied already when $N_\tau=2\rightarrow 3$, in strong contrast to
QCD or even pure gauge SU(N) phase transitions, where scaling
violations are still seen at $N_\tau = 6$.  Especially the $T=0$
quantities $g^2_R$ and $m_H/m_W$ do not show systematic $a$-dependence
beyond the statistical errors.  In this case $g_R^2 \approx 0.585$ and
$m_H/m_W \approx 0.422$, close to the tree level value.  The good
scaling behaviour indicates that the modes constant in $\tau$ are the
dominant ones.

In the physically allowed range $m_H \gsim 64$\,GeV, the inequality
$T \gg m_H(T),\,m_W(T)$ makes the 4D formalism very unwieldy.  An
appealing remedy for this is to use {\em asymmetric lattice
spacings\,} $a_s = \xi a_\tau$.  The kinetic part of
\eq\nr{lattaction} splits into temporal and spatial parts with
couplings $\beta_G^\tau$, $\beta_G^s$ and $\beta_H^\tau$, $\beta_H^s$.
The couplings are related to the asymmetry through relations
$\beta_i^\tau/\beta_i^s = f_i(\xi)$.  These have been evaluated
perturbatively to order $O(g^2,\lambda)$ by requiring isotropy in the
$W$ and Higgs correlations \cite{Csikor_Fodor_asymm}.  In
non-perturbative tests these relations were reproduced very
accurately~\cite{Csikor_asymm_sigma}.

\sectopspace
\subsection{Lattice action in 3D}\label{sec:3dlatt}

The 3D SU(2)-fundamental Higgs lattice action is similar in form to
the 4D action in \eq\nr{lattaction}, except now the indices are
limited to values 1--3.  The essential difference between the 4D and
3D formalisms becomes evident when we look at the derivation of the
CPCs.  The crucial point is the superrenormalizability of the 3D
action \nr{3daction}: $g_3^2$ and $\lambda_3$ do not run, and
$m_3^2(g_3^2)$ has only 1- and 2-loop divergences.  Calculating the
relevant diagrams both in the continuum and on the lattice one obtains
the relations~\cite{klrs_generic,klrs_nonpert}
\begin{eqnarray}
g_3^2a & = & {4\over \beta_G}  \la{3dlattcoupling}  \\
     x & = & {1\over4}\lambda_3 a \beta_G =
             {\beta_R\beta_G\over\beta_H^2} \la{xtolatt} \\
     y & = &
	{\beta_G^2\over 8} \biggl({1\over\beta_H}-3
	-{2x\beta_H\over\beta_G}\biggr)
    	+{3\Sigma\beta_G\over 32\pi}(1+4x)  \nonumber     \\
     & + &{1\over16\pi^2}\biggl[\biggl({51\over16}+9x-12x^2\biggr)
          \biggl(\ln{3\beta_G\over2}+\zeta\biggr) \nonumber \\
     & + & 5.0 + 5.2 x\biggr].	\la{ytolatt}
\end{eqnarray}
Eq.\,\nr{ytolatt} depends on several constants arising from lattice
perturbation theory: $\Sigma = 3.17591$, $\zeta = 0.09$, and the two
numbers 5.0 and 5.2, specific for SU(2)-Higgs theory and calculated in~\cite{Laine95}.

For fixed continuum parameters $(g_3^2,x,y)$ 
eqs.\,(\ref{3dlattcoupling}--\ref{ytolatt}) define the CPC in the
space of $(\beta_G,\beta_H,\beta_R)$ when lattice spacing $a$ is
varied.  Also note that in 3D $\beta_G \propto 1/a$.  In contrast to
the 4D case the continuum limit is well defined in 3D\@.  The equations
above have relative accuracy $O(a^{-1})$, so that the relation
continuum$\leftrightarrow$lattice becomes exact when $a\rightarrow 0$.
However, in practice the finite $a$ effects have been observed to be
small.

The Bielefeld group \cite{Karsch_3d80,Karsch_3d100} takes a different
philosophy to 3D effective theories: they do not utilize the
superrenormalizability of the 3D action, but consider that the most
natural approach is to fix the cutoff scale to be of the same order of
magnitude than the physical scales.  While this introduces differences
$\propto a$, the smallness of the finite $a$ effects makes most of the
results comparable to the $a\rightarrow 0$ results.

\sectopspace
\section{PHASE TRANSITION}
\secspace

\subsection{Phase diagram} 

In previous Lattice meetings~\cite{Kajantie95,Jansen96} results with
$m_H$ from 18 to 80\,GeV were reported.  The transition was seen to be
quite strongly 1st order at small $m_H$, and to weaken rapidly with
increasing $m_H$.  When $m_H=80$\,GeV the resolution was not good
enough to distinguish the order of the transition.  Since this mass
region is expected to be physically relevant, it is essential to
clarify the situation at larger $m_H$.

An important point in understanding the phase structure is the
observation that the EW theory does not have a true gauge invariant
order parameter which would distinguish the symmetric and Higgs
phases.  Indeed, it has been shown analytically that in the
SU(2)-Higgs lattice system where the Higgs length is fixed the Higgs
and the confined phases are analytically connected \cite{Fradkin79} in
the strong coupling regime; this was also observed in the early
simulations \cite{Damgaard87,Evertz87}.

A study of one-loop Schwinger-Dyson equations \cite{Buchmuller95}
argues in favour of the end of the 1st order transitions at $m_H\sim
100$\,GeV, after which only a smooth cross-over remains.  This is
certainly consistent with the observations above.  However, the result
relies on the applicability of the perturbation theory, which is known
to break down at $m_H\sim 80$\,GeV\@.  On the other hand,
$\epsilon$-expansion techniques \cite{epsilon} predict that a weak 1st
order transition remains even with large $m_H$.  Due to the lack of an
order parameter, it is not likely that the 1st order transition turns
into a line of 2nd order transitions.

\begin{figure}[tb]
\vspace{4mm}
\epsfxsize=8.2cm
\centerline{\epsfbox{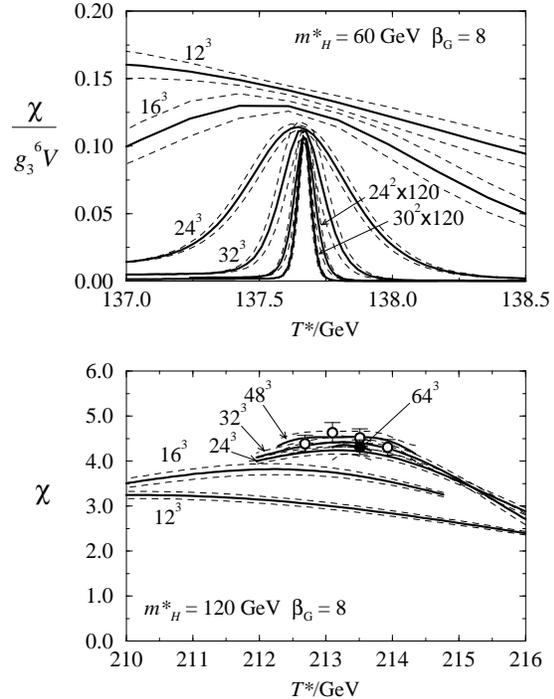}}
\vspace{-3cm}
\caption[0]{$\chi(T)$ for $m_H^*=60$ and 120\,GeV
around the maximum \cite{klrs_heavy} (Note the different
$y$-axes!).}
\label{fig60and120susc}
\vspace*{-3mm}
\end{figure}

\begin{figure}[tb]
\vspace{-8mm}
\epsfxsize=8.5cm
\centerline{\epsfbox{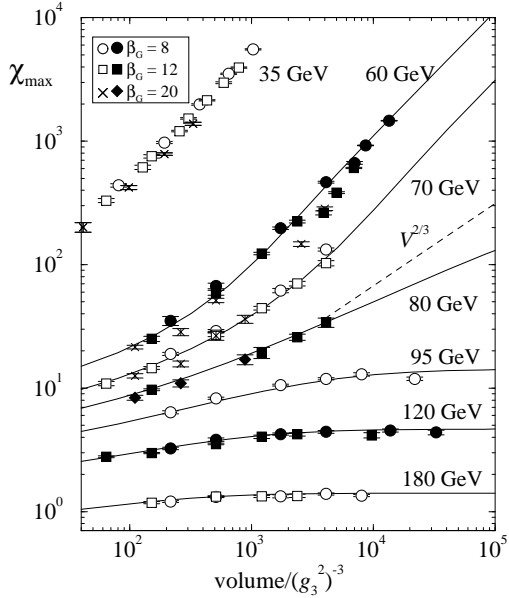}\hspace*{0.6cm}}
\vspace{-3.9cm}
\caption[0]{$\chi_{\rm max}$ for different $m_H^*$ as a
function of $V$.  The continuous lines are mean field fits, the
dashed line corresponds to the mean field critical exponent.}
\label{figchimax}
\vspace{-3mm}
\end{figure}

The situation has been addressed by the recent 3D
\cite{klrs_heavy,Karsch_3d80,Karsch_3d100} and 4D \cite{Aoki96}
simulations, which indicate that the transition turns into a smooth
cross-over at $m_H\approx 80$\,GeV\@.  Kajantie \etal
\cite{klrs_heavy} utilized the finite size behaviour of the
$\phi^\dagger\phi$ susceptibility (in 3D notation):
\be
  \chi = g_3^2 V \langle ( \phi^\dagger\phi - 
	\langle \phi^\dagger\phi \rangle)^2 \rangle
\ee
where $ \phisq/g_3^2 = \beta_G\beta_H \langle R^2\rangle/8 + const.$
and $V(g_3^2)^3 = (4N/\beta_G)^3$.  For each $V$ the temperature is
adjusted until the maximum value of $\chi$ is found.  According to the
finite size scaling, in 1st order transitions $\chi_{\rm max} \propto
V$, in 2nd order transitions $\chi_{\rm max} \propto V^{\gamma/3}$,
where $\gamma$ is a critical exponent, and if there is no transition
$\chi_{\rm max}$ approaches a constant value.  In
\fig{fig60and120susc} $\chi(T)$ is shown for $m_H^*=60$ and
120\,GeV\@.  The difference is striking: the quantities $\chi_{\rm
max}/V$ (60\,GeV) and $\chi$ (120\,GeV) approach constant values,
consistent with a 1st order transition and no transition,
respectively.  $\chi_{\rm max}$ for $m_H^* = 35$--180\,GeV are plotted
against $V$ in \fig{figchimax}.  Scaling is well satisfied: points
with different $\beta_G$ fall on the same curve.  To emphasize the
approach to the asymptotic $V^1$, $V^0$ -lines, a simple mean field
model has been fitted to the data; the results of the fits are
shown as continuous curves.

Similar behaviour was reported by Y.\,Aoki \cite{Aoki96}, using 4D
$m_H=47$--102\,GeV simulations, and by the Bielefeld group
\cite{Karsch_3d100}, using $m_H=60$--100\,GeV\@ in 3D.  Interestingly,
also in finite $T$ U(1)-Higgs theory an endpoint of the 1st order
phase transitions has been observed~\cite{Karjalainen96}.

A more detailed study of the properties of the critical point was
performed in \cite{Karsch_3d100}.  In the $m_H=80$\,GeV case $m_H(T)$ was
measured around the transition in the symmetric and broken phases and
fitted to the ansatz
\be
  m_H \propto |\beta_G - \beta_{G,c}|^{\nu}
\ee
(here $\beta_G$ is adjusted while keeping $\beta_H$, $\beta_R$
constant), with the result $\nu = 0.49(2)$ and 0.31(1) in the
symmetric and broken phases, respectively.  The different indices
could indicate a tricritical nature for the endpoint; on the other
hand, 80\,GeV is likely not the exact value of $m_{H,c}$.  Indeed,
utilizing the analysis of the Lee-Yang zeroes for $m_H=80$
and 100\,GeV\@ it was estimated that $m_{H,c} \approx 77$\,GeV.

\sectopspace
\subsection{$T_c$ and metastability}

\begin{figure}[tb]
\vspace*{-5mm}
\epsfxsize=7cm
\centerline{\hspace{-5mm}\epsfbox{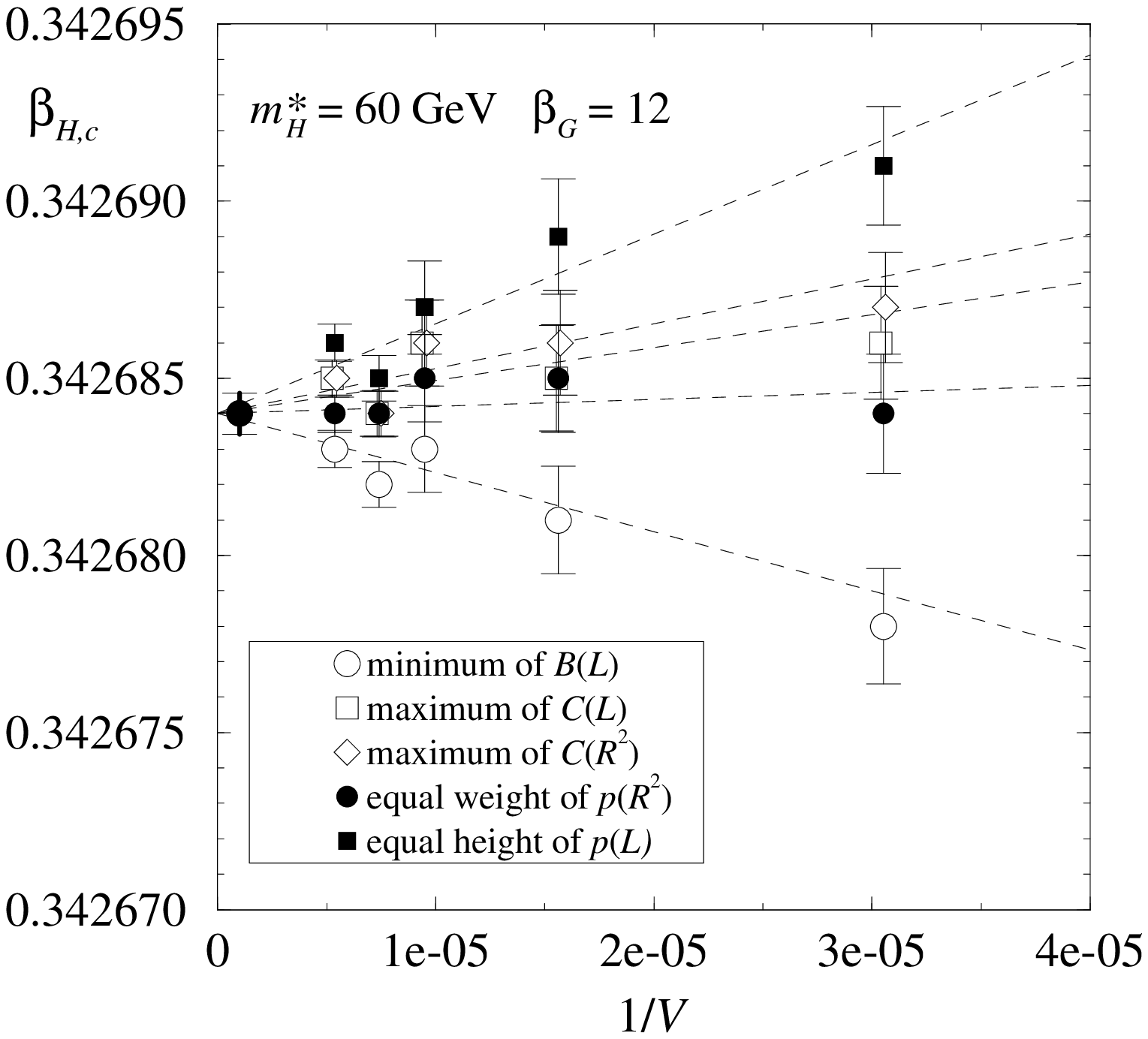}}
\vspace*{-4cm}
\epsfxsize=7cm
\centerline{\hspace{-5mm}\epsfbox{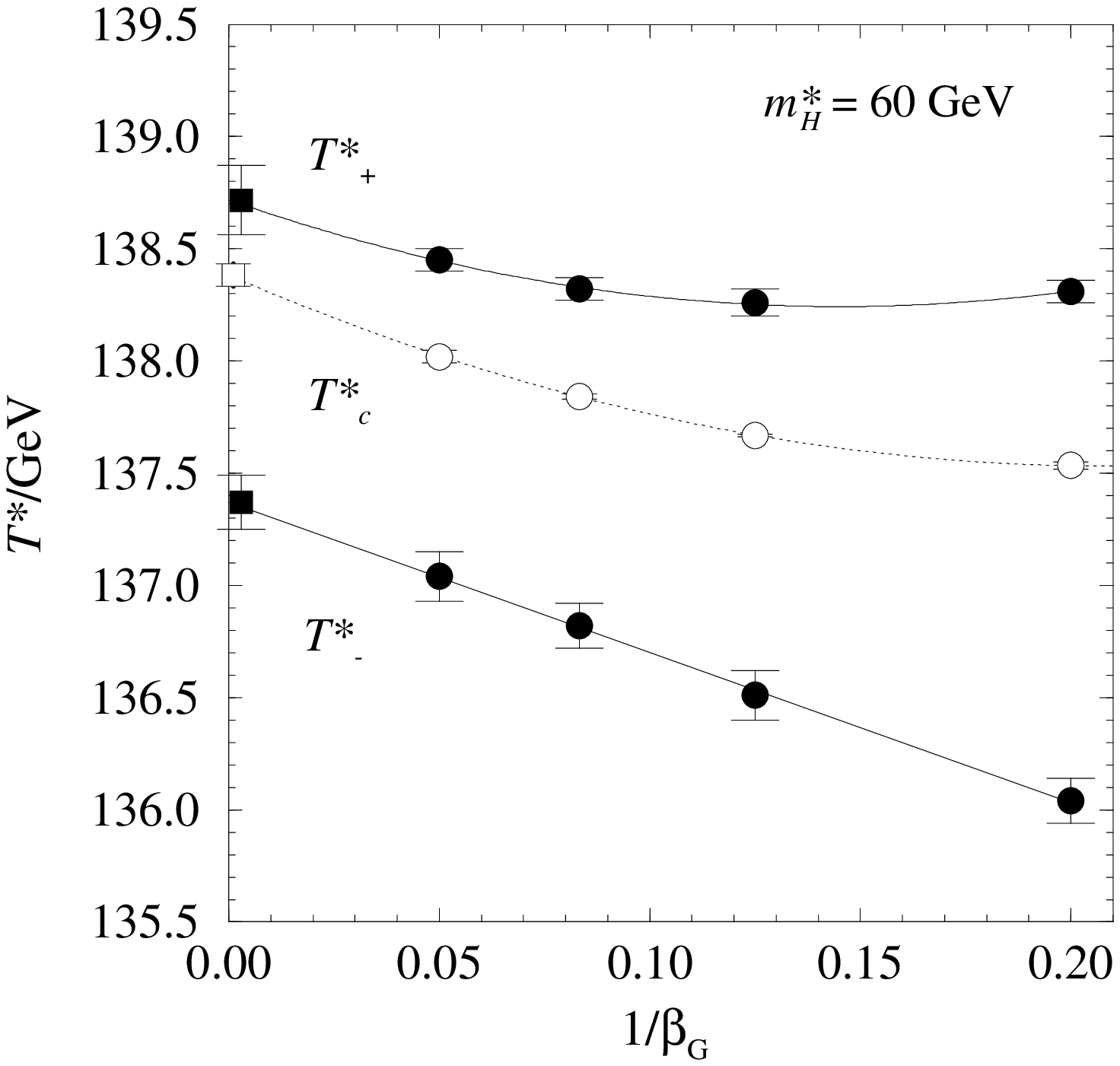}}
\vspace*{-4.4cm}
\caption[0]{Top: the $V\rightarrow\infty$ extrapolation of the
pseudocritical coupling $\beta_{H,c}$.  Bottom: the continuum limit
($1/\beta_G=0$) extrapolation of $T_c$ (open symbols).  Also shown
here is the metastability range (filled symbols) \cite{klrs_nonpert}.}
\label{figm60tc}
\vspace*{-3mm}
\end{figure}

To accurately locate $T_c$ both the finite size and finite lattice
spacing effects have to be addressed.  In 3D simulations, for fixed
$\beta_G = 4/(g_3^2a)$, $x$ and $V$, one adjusts $y$ in order to find
the pseudocritical point ($\beta_H$ and $\beta_R$ are given by
eqs.\,(\ref{xtolatt}--\ref{ytolatt})).  Through
eqs.\,(\ref{g32}--\ref{y}) this corresponds exactly to adjusting $T^*$
while keeping $m_H^*$ constant.

With finite $V$, there are several non-equivalent methods to locate
the pseudocritical point: the maximum of the order parameter
susceptibility, the minimum of the Binder cumulant and the ``equal
weight'' and ``equal height'' points of the order parameter
distributions \cite{klrs_nonpert,Gurtler96}.  In the
$V\rightarrow\infty$ limit these all extrapolate to the same value, as
shown in top part of \fig{figm60tc}.  Multicanonical simulations and
histogram reweighting are commonly used in the analysis.
This is repeated for 2--4 values of $\beta_G$ while $x$ is kept
constant.  The continuum limit is obtained by extrapolating in
$1/\beta_G$.  This is shown in \fig{figm60tc} for $m_H^*=60$\,GeV; in
this case the range in $\beta_G$-values (5--20) is so large that
the subleading behaviour is seen.  Nevertheless, it should be noted
that the $T^*$ varies very little across the extrapolation: the
curvature is seen only because of the very small statistical errors.

The phase {\em metastability range\,} is also shown in \fig{figm60tc}.
This has been obtained as follows: at $y_c$, the order parameter
histograms have a 2-peak structure.  The histograms are
reweighted off $y_c$ until the ``shoulder'' of one of the peaks
vanishes; in terms of the constrained effective potential this
corresponds to the temperature at which the metastability of the phase
vanishes~\cite{klrs_nonpert}.

The 3D effective theory describes a whole class of 4D theories, as
discussed in sec.\,\ref{sec:3daction}.  In table \ref{tab:tc} the
physical $T_c$ is shown for 4D SU(2)+Higgs and SU(2)+Higgs+fermions
-theories, in addition to the ``bare'' $T_c^*$ -values.  The fermion
content is the same as in the MSM, with $m_{\rm top} = 175$\,GeV\@.
Note that the physical $m_H$ is different from $m_H^*$, and that the
$m_H^*=35$\,GeV -set does not correspond to any physical fermion
theory.

\begin{table}[bt]
\caption[0]{$T^*_c$ and $T_c$ for physical SU(2)+Higgs and physical
SU(2)+Higgs+fermions -theories.  $m_H^*=35$--70 is from \cite{klrs_nonpert},
72.18 from \cite{Gurtler96} (70\,GeV in authors' notation).  All units
are in GeV.}
\label{tab:tc}
\begin{tabular*}{\columnwidth}{@{}l@{\extracolsep{\fill}}lllll}
\hline
$m_H^*$        & 35         & \n60       &  \n70       &  \n72.18   \\
\hline
$T_c^*$        & 92.64(7)   & 138.38(5)  &  154.5(1)   &  157.74(5) \\
$T_{c,\rm pert}^*$& 93.3    & 140.3      &  157.2      &  160.9     \\
\hline
\multicolumn{5}{l}{SU(2)+Higgs} \\
$m_H$          & 29.1       &\n54.4      &\n64.3       & \n66.5     \\
$T_c$          & 76.8       & 132.6      & 151.2       &  154.7     \\
\hline
\multicolumn{5}{l}{SU(2)+Higgs+fermions ($m_{\rm top} = 175$\,GeV)} \\
$m_H$          & --         &\n51.2      &\n68.0       & \n69.4     \\
$T_c$          & --         &\n89.8      & 105.8       &  107.2     \\
\hline
\end{tabular*}
\vspace*{-8mm}
\end{table}

\begin{table}[bt]
\caption[0]{$T_c$ from 4D SU(2)-Higgs simulations.}\label{tab:tc4d}
\begin{tabular*}{\columnwidth}{@{}l@{\extracolsep{\fill}}llll}
\hline
$m_H$/GeV & $T_c/m_W$ & $N_\tau$ &  \\
\hline
  16      & 0.464(2)  &   3      & \mbox{ \cite{desy_18_49}}\\
  34      & 0.910(16) & $\infty$ & \mbox{ \cite{desy_34}}   \\
  48      & 1.153(16) &   3      & \mbox{ \cite{desy_18_49}}\\
\hline
\end{tabular*}
\vspace*{-4mm}
\end{table}

Locating the critical couplings in 4D simulations is essential for
determining CPC, as discussed in sec.\,\ref{sec:4daction}.  The
methods above can be used also in 4D to locate the critical couplings;
in addition the DESY group has used methods which rely
on the coexistence of two phases in long cylindical lattices
\cite{desy_18_49,desy_34}: in the ``constrained'' method the order
parameter is restricted to a narrow range between the pure phase
values, enforcing the system to reside in a mixed phase.  The coupling
is then tuned so that the distribution in this region is
horizontal. This is equivalent to the condition that the `flat' part
of the 2-peak histogram is horizontal, but requires much less cpu-time
than the full histogram calculation.  In the ``2-coupling'' method the
system is split into 2 halves, and the critical coupling is bracketed
by tuning the couplings in the two subvolumes individually so that the
2-phase configuration is maintained.

In table \ref{tab:tc4d} the ratio $T_c/m_W$ is shown for different $m_H$.
For the 34\,GeV case results from $N_\tau=2$--5 have been used to 
extrapolate $a\rightarrow 0$ quadratically.

\begin{figure}[tb]
\figtopspace
\epsfxsize=\figfxsize
\centerline{\epsfbox{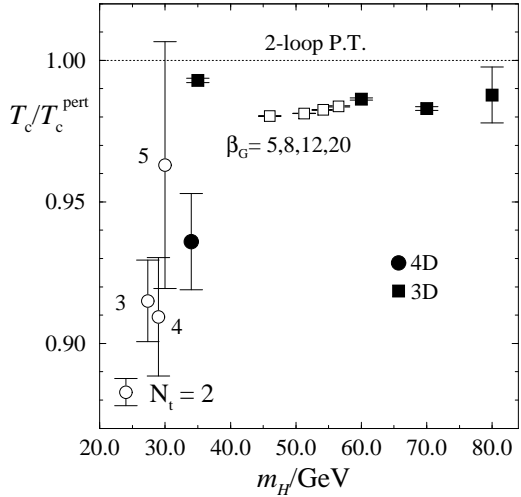}}
\figbottomspace
\caption[0]{$T_c/T_c^{\rm pert}$ from 4D \cite{desy_34} and
3D \cite{klrs_nonpert} simulations.  The open symbols show the
extrapolation to the continuum limit (only $m^*_H=60$ for 3D) and
have been shifted horizontally for clarity.}
\label{figtc}
\vspace*{-3mm}
\end{figure}

A direct comparison between 3D and 4D simulations at this stage is not
straightforward, due to very different connections to continuum
physics.  The 3D simulations have used $g^2(7T) \approx 0.444$,
whereas in 4D simulations $g_R^2 \approx 0.58$.  The relations
$(g_3^2,x,y)\leftrightarrow $\,4D quantities allows one to adjust
$g^2$; however, then also the physical $m_H$ changes.  Currently there
are no simulations which would correspond to the same physical
situation (however, see \cite{Laine96}).  Nevertheless, it is
straightforward to compare the results individually to 2-loop
perturbative results.  In \fig{figtc} $T_c/T_c^{\rm pert}$ is plotted,
the squares correspond to 3D $m^*_H=35$, 60, 70 and 80\,GeV cases,
circles to 4D $m_H=34$\,GeV\@.  Also shown is the approach to the
continuum limit; for 3D the points are the same as shown in
\fig{figm60tc}.  A striking feature is the smallness of the
statistical errors in the 3D simulations.  This is due to the rigorous
nature of CPCs in 3D: the errors in $T_c$ are directly translated from
the statistical errors of critical couplings.  In 4D the errors
accumulate through $T=0$ mass measurements.  For the bare values of
the critical couplings the accuracy is comparable, up to 6--7 decimal
places.  The finite $a$ effects are seen to be small in both cases,
and $T_c/T_c^{\rm pert}$ is consistently smaller than 1.  Given the
difference in the values (and renormalization schemas!) of $g^2$ and
the 4D statistical errors the results can not be considered
inconsistent.

\sectopspace
\subsection{Interface tension and latent heat}

\begin{figure}[tb]
\vspace{-1cm}
\epsfxsize=\figfxsize
\centerline{\hspace{-4mm}\epsfbox{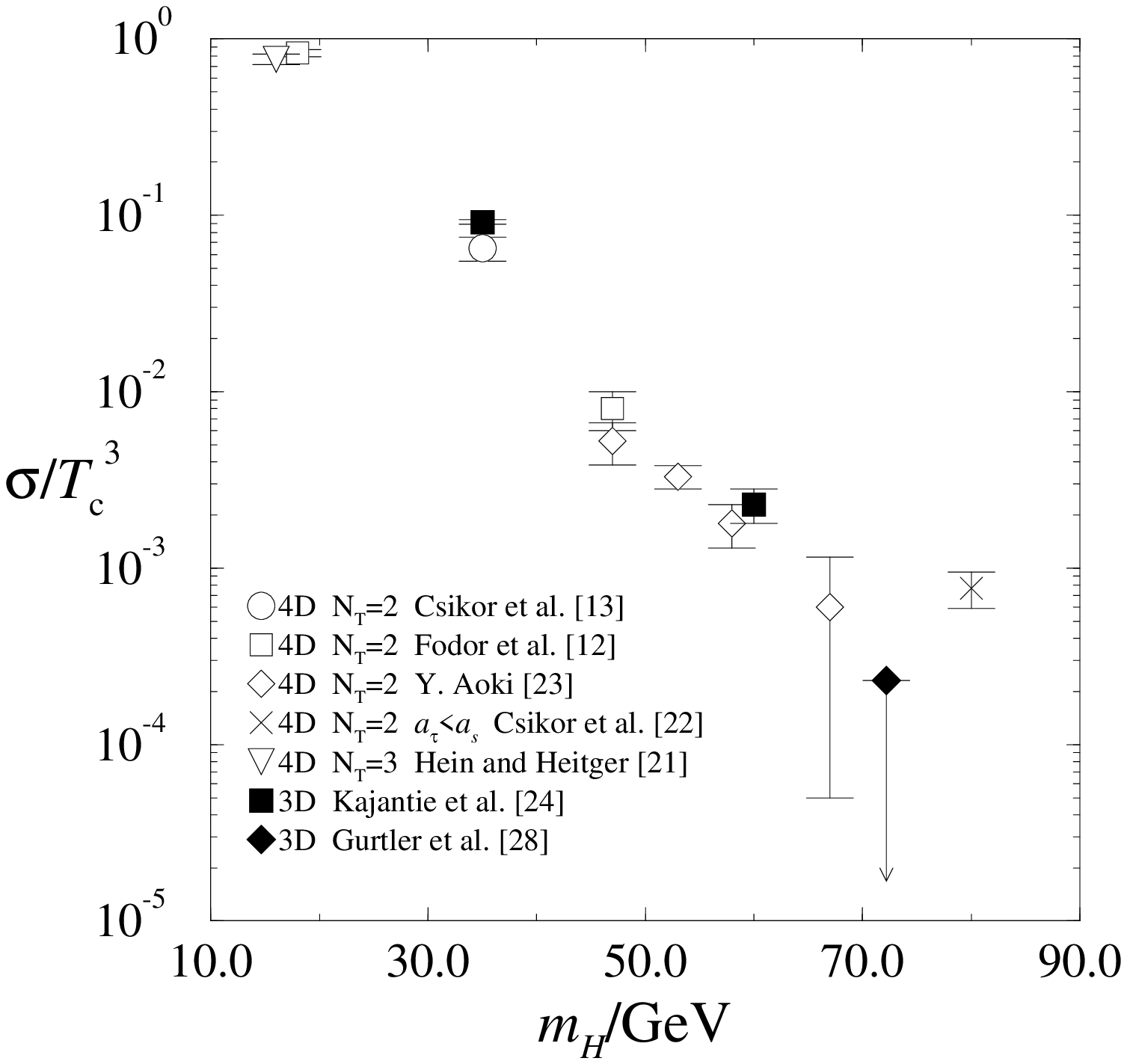}}
\vspace*{-5cm}
\epsfxsize=\figfxsize
\centerline{\hspace{-4mm}\epsfbox{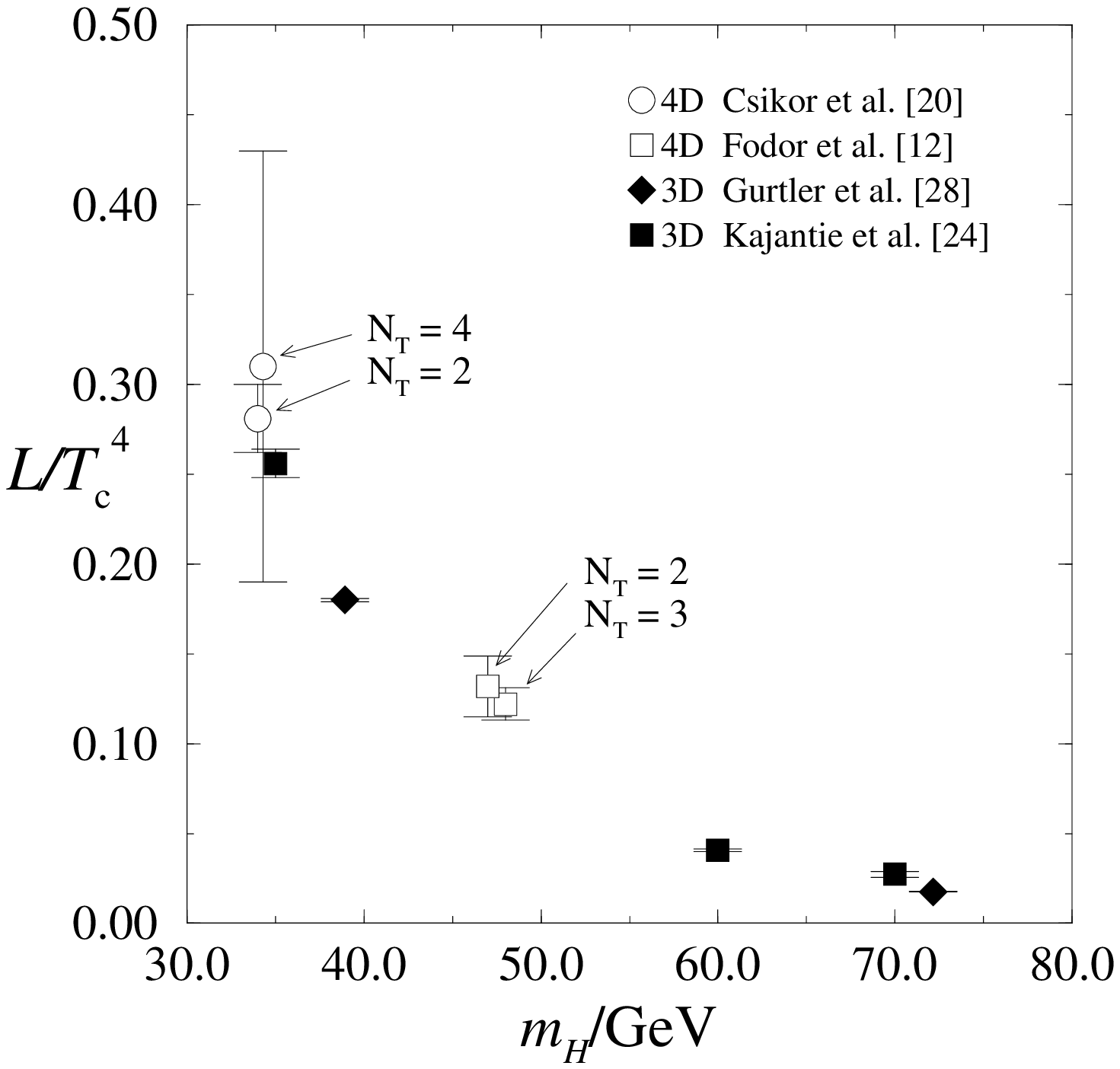}}
\figbottomspace
\caption[0]{The interface tension (top) and latent heat (bottom) from
	4D and 3D simulations.}
\label{figsigmaL}
\vspace*{-3mm}
\end{figure}

The tension of the interface $\sigma$ between the 2 coexisting phases
and the latent heat $L$ of the transition are primary quantities
characterizing the strength of the transition.  The now ubiquitous
{\em histogram method\,} \cite{Binder82} is the most accurate but
numerically very demanding way to measure $\sigma$: for each lattice
3-volume $V = a^3 L^2 L_z$, where $L \le L_z$ is assumed, one measures
$p_{\rm max}$ and $p_{\rm min}$, the probability distribution maximum
and the minimum between the peaks at $T_c$.  $\sigma$ is now obtained
at the $V\rightarrow\infty$ limit of the expression
\be
\fr{\sigma a^2}{T} = \fr1{2L_x^2} \bigg[ 
	\ln\fr{p_{\rm max}}{p_{\rm min}}
	+ \fr12\ln\fr{L^{3}_z}{L^2_x} + G + c\bigg]
\la{sigma}
\ee
where $G=\ln 3$ for $L_x=L_z$ and 0 for $L_x\ll L_z$, and c is a
constant.  In addition, $a\rightarrow 0$ limit should be taken.
Eq.\,\nr{sigma} has been used both in 3D and 4D simulations
\cite{desy_18_49,klrs_nonpert,Gurtler96,Aoki96}.  The DESY
group \cite{desy_18_49,desy_35interface,Hein_16,Csikor_asymm_sigma}
has also used the {\em 2-coupling\,} integration method
\cite{Potvin89} and the {\em tunnelling correlation length\,}
\cite{Jansen88} analysis.  These methods are not quite as demanding
computationally as the histogram method.

In \fig{figsigmaL} $\sigma$ is shown from several calculations.  Both
the 3D calculations shown perform $a\rightarrow 0$ extrapolation.
Note the dramatic decrease in $\sigma$ when $m_H$ increases.
Considering the difficulties in measuring $\sigma$ reliably the
agreement must be considered to be good; the only point which
disagrees somewhat with the trend is the 80\,GeV asymmetric lattice
($a_\tau = a_s/4$, $N_\tau=2$) point presented in this conference
\cite{Csikor_asymm_sigma}.  Perturbatively, $\sigma$ can be calculated
with any reliability only when the transition is strong: indeed, at
very small $m_H$ the agreement between the lattice and perturbative results
is fair, but already at $\sim 60$\,GeV the lattice results are a factor of
4--5 smaller.

The latent heat $L$ is directly related to the jump of $\phisq$ at the
transition through the Clausius-Clapeyron equation:
\be
  L/T_c^4 = \fr{m^2_H}{T_c^3}  \Delta \phisq.
\la{latent}
\ee
In \fig{figsigmaL} $L$ is shown from 3D and 4D simulations; in 3D the
extrapolation to the continuum limit has been performed.  Both in 3D
and 4D also alternative methods to \eq\nr{latent} have been used for
measuring $L$, with practically unchanged results.  Within errors, the
results are consistent with the 2-loop perturbation theory.

\sectopspace
\subsection{Interaction measure}

The interaction measure $\delta \equiv \epsilon - 3p$ characterizes
the deviation of the system from the massless ideal gas behaviour.
This has been measured by the DESY group in 4D at $m_H=34$\,GeV
\cite{desy_34}; in 3D $\delta$, $\epsilon$ and $p$ are not readily
accessible.  When $0.5 < T/T_c \le 1$ in the broken phase, $\delta/T^4
\approx 0.6$, and at $T_c$ it jumps discontinuously to $\approx 0.9$.
When $T$ increases further $\delta/T^4$ falls rapidly, and at
$T\approx 2T_c$ it is $\sim 0$.  This qualitatively agrees with the
perturbation theory in the broken phase, where it is applicable.

\sectopspace
\section{BROKEN AND SYMMETRIC PHASE}
\secspace

Deep in the broken phase perturbative analysis is well controlled.
Lattice studies at $T \lsim T_c$ can reveal the accuracy and the
eventual failure of the perturbation theory.  Indeed, as mentioned in
the introduction, the expansion parameter is $\sim g_3^2/(\pi m_W)$,
where $m_W(v) \rightarrow 0$ formally when the symmetry is restored.

In refs.\,\cite{klrs_nonpert,Gurtler96} the behaviour of $\phisq$ is
studied with 3D simulations.  The general agreement between the
lattice and continuum 2-loop perturbative results is good,
progressively becoming worse when $T_c$ is approached.  However, since
$\phisq$ can be determined very accurately in the broken phase
(relative error $\sim 10^{-3}$), deviations are seen even relatively
deep in the broken phase.  This information was used in
\cite{klrs_nonpert} to infer the value of the so far uncomputed 3-loop
term in the effective potential, and to verify that it is linear in
$\phi$.  Even with this correction the effective potential fails at
the transition point.

\begin{figure}[tb]
\vspace{-2.8cm}
\epsfxsize=6.5cm
\centerline{\epsfbox{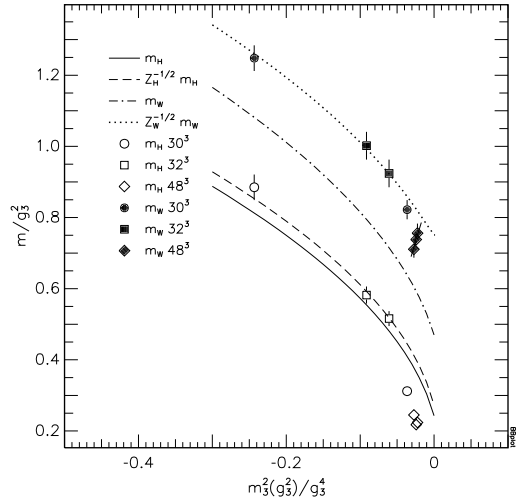}}
\vspace{-0.5cm}
\caption[0]{$m_H$ and $m_W$ in the broken phase at $m_H^*=70$\,GeV
compared to the perturbation theory (G\"urtler \etal
\cite{Gurtler96}).}
\label{figbrokenphase}
\vspace*{-3mm}
\end{figure}

The Higgs and W screening masses are measured with the scalar ($0^{++}$) and
vector ($1^{--}$) operators of type
\ba
   S_x       &=& \tr [\Phi_x^\dagger\Phi_x]  \la{scalarop} \\
   V^a_{x;i} &=& 
\tr [\tau^a\Phi_x^\dagger U_{x;i}\Phi_{x+\hat\imath}]
\la{vectorop}
\ea
where $\tau^a$ is a Pauli matrix.  The masses (inverse correlation
lengths) are extracted from the correlation functions $\langle S_x S_y
\rangle$ and $\langle V_x V_y \rangle$; in the $V$-correlations only
the diagonal part survives.  To improve the signal the operators can
be smeared or blocked in various ways.  

In \fig{figbrokenphase} the measured values of $m_H$ and $m_W$ are
compared against 1-loop perturbative values (with and without the wave
function renormalization) for $m_H^*=70$\,GeV\@ \cite{Gurtler96}.
Deep in the broken phase the results agree within errors;
however, closer to the transition deviations appear.
Note that perturbatively the transition occurs at $m_3^2=0$, whereas
in the actual simulations $m_{3,\rm c}^2 < 0$.  Similar behaviour has been
observed in other simulations~\cite{klrs_nonpert,Karsch_3d80,desy_18_49}.

\begin{figure}[tb]
\epsfxsize=7cm
\centerline{\epsfbox{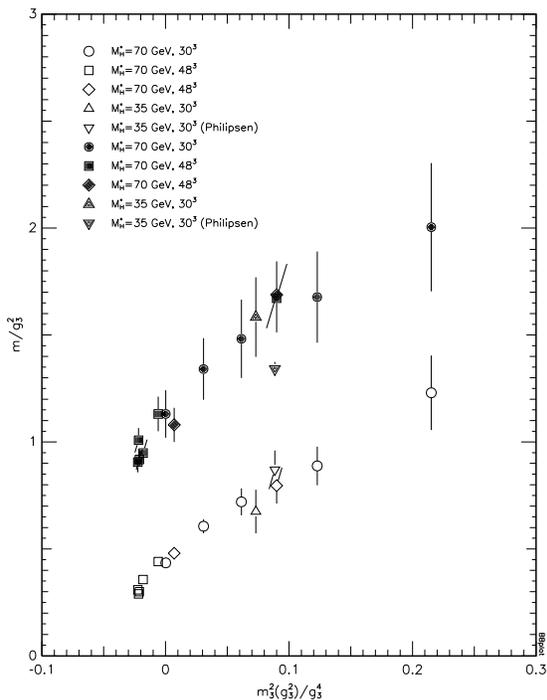}}
\vspace{-1.6cm}
\caption[0]{Vector (filled) and scalar (open) masses in the symmetric
phase \cite{Gurtler96}.}
\label{figsymmphase}
\vspace*{-3mm}
\end{figure}

The symmetric phase vector and scalar masses, measured with operators
(\ref{scalarop}-\ref{vectorop}), are shown in \fig{figsymmphase}
\cite{Gurtler96}.  Both masses increase when $T$ ($m_3^2$) increases.
An interesting observation is that the $m_H^*=35$ and 70\,GeV masses
are equal -- this is reasonable, because in the symmetric phase the
$\lambda(\phi^\dagger\phi)^2$ -term in the action \nr{3daction} is
very small.  Comparing Figs.\,\ref{figbrokenphase} and
\ref{figsymmphase} we note that there is a small but clearly
discernible discontinuity of the masses at the transition: in the
symmetric phase the masses become larger.  Again, similar behaviour
has been observed by other groups~\cite{klrs_nonpert,Karsch_3d80}.

Philipsen \etal \cite{Philipsen96} used various levels of blocking of
the operators (\ref{scalarop}-\ref{vectorop}), and measured the full
correlation matrix between different blocking levels.  By performing
an eigenstate analysis the ground state and a few lowest exited states
could be distinguished.  The ground states in the vector and scalar
channels are shown in \fig{figsymmphase} with downwards pointing
triangles.  In the scalar channel the mass agrees with the
measurements of \cite{Gurtler96}, but the vector mass is slightly
smaller, likely due to the better projection to the ground state.  In
the scalar channel also standard plaquette `$W$-ball' operators are
used together with the blocked $S$-operators.  In the symmetric phase
very little mixing is seen between these operators; furthermore, the
masses in the $W$-ball sector are observed to be almost identical to
the 3D SU(2) $0^{++}$ glueball masses measured at the same value of
$\beta_G$~\cite{Teper92}.

\begin{figure}[tb]
\vspace*{-8.6cm}
\epsfxsize=9.5cm
\centerline{\hspace{3cm}\epsfbox{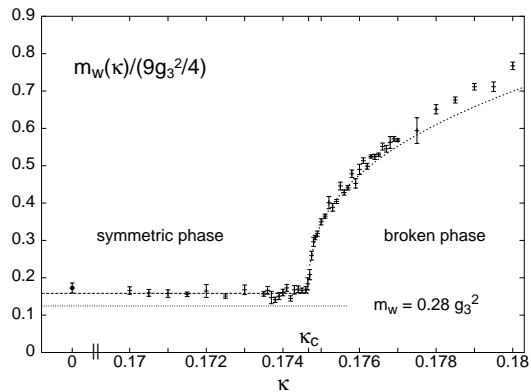}}
\vspace{-5mm}
\caption[0]{$m_W$ in the Landau gauge (after \cite{Karsch_3d80}).}
\label{figgaugefix}
\vspace*{-3mm}
\end{figure}

In \cite{Karsch_3d80} $m_W$ was measured by fixing to the Landau gauge
and using $A_i^a$ -operators to measure the correlations.  The results
are shown in \fig{figgaugefix}: in the broken phase, the results are
equal to those measured with the operators of type
\eq\nr{vectorop}.  However, in the symmetric phase $m_W \approx 0.35
g_3^2$, independent of $m_3^2$ (and $T$), in strong contrast to the
$V$-operator in \fig{figsymmphase}.  The point at $\kappa=0$ in
\fig{figgaugefix} is a pure gauge result, i.e.\@ calculated without
the Higgs field.

Can one understand this behaviour analytically?  In the symmetric
phase the perturbative expansion breaks down.  The system may
essentially behave like 2+1D SU(2) gauge theory at $T=0$, which is
confining.  The Higgs field can be interpreted as a scalar quark, and
the physical states are now $W$-balls and $(\phi^\dagger\phi)$ bound
states.  Using this strong coupling picture and the static
$\phi$-$\phi$ potential determined with lattice simulations
\cite{Ilgenfritz95}, Dosch \etal \cite{Dosch95} calculated the mass
spectrum of the bound states analytically, reproducing the pattern in
\fig{figsymmphase} quite well.  On the other hand, the 1-loop
Schwinger-Dyson gap equation calculation by Buchm\"uller and Philipsen
\cite{Buchmuller95} gives an approximate result $m_W \approx
0.28\,g_3^2$ in the symmetric phase.  This is obviously in conflict
with the vector masses in \fig{figsymmphase}, but is close to the
Landau gauge fixed result, and is shown as a horizontal dashed line in
\fig{figgaugefix}.

The static potential and the string tension $\sigma$ in the symmetric
phase have been measured in \cite{Ilgenfritz95,Gurtler96}.  Also in
this case no significant dependence of $\sigma/g_3^4$ on $\lambda$
($m_H^*$) was observed.  When $m_3^2$ ($T$) is large, the value of
$\sigma$ was close to the pure 3D SU(2) gauge theory value
\cite{Teper92}.  At large distances one expects the screening
behaviour to set in; however, this has not been observed yet in the
distances presently allowed by the available resources.

These results, while clearly supporting the confinement picture in the
symmetric phase, also show that the gauge degrees of freedom decouple
almost completely from the Higgs field.  Indeed, all the measurements
which involve only gauge fields (plaquette correlators, Landau gauge
$A_i^a$ correlator, string tension) give results almost identical to
the pure SU(2) gauge theory.

\sectopspace
\section{SPHALERON TRANSITION RATE}
\secspace

Due to the axial anomaly the baryon number is not conserved in the EW
theory.  In order to quantify the effects of the EW physics to the
baryon number of the Universe \cite{baryogenesis} the knowledge of the
sphaleron transition rate $\Gamma$ is essential.  In terms of the
gauge fields, $\Gamma$ is the diffusion rate of the topological charge
$B(t) = \Delta N_{\rm CS}(t)$ where $N_{\rm CS}$ is the Chern-Simons
number.  The effective potential is periodic to the $N_{\rm CS}$
-direction: there are large gauge transformations which change $N_{\rm
CS}$ by unity.  The goal is to calculate the rate of the dynamical
processes which change $N_{\rm CS}$, driving the configuration from
one minimum to a neighbouring one.  Since $B(t)$ can be described as a
random walk in the periodic potential,
\be
  \langle B^2(t)\rangle_T \rightarrow \Gamma Vt, \h t\rightarrow\infty
\ee
where $\langle\cdot\rangle_T$ is the canonical expectation value
\cite{Khlebnikov88}.  From analytical considerations \cite{Arnold87}
one expects $\Gamma = \kappa (\alpha_W T)^4$, where
\be
\begin{array}{llll}
  \kappa & = & \mbox{const.} & T > T_c \\
  \kappa & \approx & f \exp [-E_{\rm sph}(T)/T] & T < T_c
\end{array}
\la{rate}
\ee
The calculation of $\Gamma$ is non-perturbative.  Since $\Gamma$ is a
real-time transition rate, the standard imaginary time lattice
formalism is not easily applicable.  However, one expects that at high
$T$ the transitions occur predominantly through thermally activated
classical processes.  This suggests the following strategy: $B(t)$ is
calculated by solving the {\em classical\,} equations of motion, and
the results are averaged over a canonical ensemble \cite{Grigoriev89}.
It is important that the Gauss constraint is satisfied.  This is
simple in the 1+1D U(1)-Higgs model, which has been studied in detail
as a prototype model \cite{Shaposhnikov92,Bochkarev94}, but for 3+1D
SU(2) correct methods have been derived only recently
\cite{Krasnitz95,Moore96}.

\begin{figure}[tb]
\epsfxsize=6.2cm
\centerline{\epsfbox{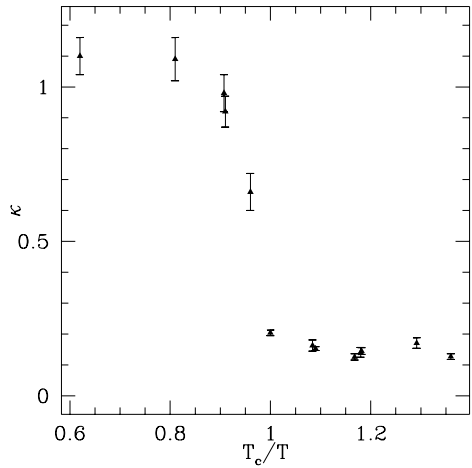}}
\vspace{-1cm}
\caption[0]{$\Gamma/(\alpha_W T)^4$ for SU(2)-Higgs \cite{Tang96}.}
\label{figkappa}
\vspace*{-3mm}
\end{figure}

In the 1+1D U(1)-Higgs model the numerical and analytical results
agree (however, in the high $T$ phase some lattice spacing dependence
remains, see \cite{Smit95}).  Ambj{\o}rn and Krasnitz \cite{Ambjorn95}
measured $\Gamma$ in the high $T$ phase of the pure gauge SU(2)
theory, with the result $\Gamma=\kappa(\alpha_WT)^4$, $\kappa =
1.09\pm 0.04$.  The result was seen to be independent of (large) $V$
and (small) $a$, indicating that it survives to the continuum
limit.  This is supported by the result by Moore \cite{Moore96}, who
also measured the linear response $\Gamma_\mu$ of $N_{\rm CS}$ to a
chemical potential, $\Gamma_\mu = 2 \Gamma$.

In the 3+1D SU(2)-Higgs model it has been analytically estimated that
$\kappa\approx 0.01$ in the high $T$ phase \cite{Philipsen95}.  Tang
and Smit have recently calculated $\Gamma$ in this model
\cite{Tang96}; the results are shown in \fig{figkappa}.  In the high
$T$ phase $\kappa\approx 1$, which agrees with the SU(2) result but is
in strong contrast to the analytical estimate.  At $T_c$ $\kappa$
decreases sharply, and remains 0.1--0.2 when $T<T_c$.  The
disagreement with the analytic form in \eq\nr{rate} is dramatic: no
exponential suppression is seen, and the rate is $10^3$--$10^8$ times
larger than expected!  Similar behaviour has been observed by
Ambj{\o}rn and Krasnitz \cite{Ambjorn96}.  The success of the 1+1D
U(1)-Higgs model makes this conflict even more striking, and obviously
the reason for this must be understood before we can have trust in the
results.

\sectopspace
\section{CONCLUSIONS}
\secspace

The numerical EW simulations have been very successful: to a large
extent, the static thermodynamic properties of the EW phase transition
have been ``solved''.  The transition is strongly 1st order at small
$m_H$, becomes rapidly weaker with increasing $m_H$ and at $m_H
\approx 80$\,GeV (for SU(2)-Higgs) the line of transition ends at a
critical point, after which only a smooth cross-over remains.  For
practical purposes, the accuracy is good enough for most of the static
quantities relevant to the transition.  The properties of the critical
point itself (exact location, exponents) are not yet so well known,
this being arguably the most difficult point in the phase diagram.
These results rule out the MSM baryogenesis: it is not any more
possible that the transition is strong enough to produce the
observed $B$ asymmetry \cite{klrs_nonpert}.

The 2-loop perturbation theory yields a good guideline for the
transition at $m_H \leq 70$\,GeV, although deviations are clearly
seen.  The 3D effective theories and their accuracy are now fully
understood theoretically, and the good general agreement with the 4D
simulations is very encouraging.  However, more comparisons should be
done in order to fully quantify the accuracy: this is especially
important since the 3D theory provides a method for investigating
realistic EW (+ beyond) theories without any of the problems usually
caused by fermions, chiral or not.

There are still unanswered questions with the current results, which
must be addressed: the role of the different correlation lengths in
the symmetric phase is still not fully clarified, and the
contradiction between the analytical and numerical results for the
sphaleron rate must be understood.

\sectopspace
\subsection*{Acknowledgement}

Discussions with K. Kajantie, M. Laine, M. Shaposhnikov, K. Jansen,
Z. Fodor, K. Jansen, T. Neuhaus and O. Philipsen are gratefully
acknowledged.  This work has been partly supported by the DOE grant
FG02-91ER40661.

\end{document}